\documentclass[aps,twocolumn,prl,bibnotes,groupedaddress]{revtex4}
\usepackage{graphicx}

\begin{document}

\title[]{\sc Spin separation in cyclotron motion}

\author{L.~P. Rokhinson}
\email[]{leonid@physics.purdue.edu}
\author{V. Larkina}
\affiliation{Department of Physics, Purdue University, West
Lafayette, IN 47907 USA}

\author{Y.~B. Lyanda-Geller}
\affiliation{Naval Research Laboratory, Washington, DC 20375, USA}

\author{L.~N. Pfeiffer}
\author{K.~W. West}
\affiliation{Bell Laboratories, Lucent Technologies, Murray Hill,
New Jersey 07974 USA}

\begin{abstract}
Charged carriers with different spin states are spatially
separated in a two-dimensional hole gas. Due to strong spin-orbit
interaction holes at the Fermi energy have different momenta for
two possible spin states travelling in the same direction and,
correspondingly, different cyclotron orbits in a weak magnetic
field. Two point contacts, acting as a monochromatic source of
ballistic holes and a narrow detector in the magnetic focusing
geometry are demonstrated to work as a tunable spin filter.
\end{abstract}


\maketitle

The ability to manipulate the spin of charge carriers in a
controllable fashion is a central issue in the rapidly developing
field of spintronics\cite{wolf01}, as well as in the development
of spin-based devices for quantum information
processing\cite{divincenzo95}. Electrical injection of
spin-polarized currents has proven to be a formidable challenge.
To date, spin polarized currents have been  generated by using
either ferromagnetic materials as
injectors\cite{fiederling99,ohno99,jedema02,hammar02},  or by
exploiting the large spin splitting of electron energy levels in
strong magnetic fields\cite{folk03,hanson03}. Here we realize a
solid-state analog of the Stern-Gerlach experiment in atomic
physics\cite{gerlach22}, with spin-orbit interactions playing the
role of the gradient of magnetic field. We achieve spatial
separation of spins and bipolar spin filtering using cyclotron
motion in a weak magnetic field.

\begin{figure}[t]
\def\ffile{fig1}
\centering\includegraphics[scale=1.1]{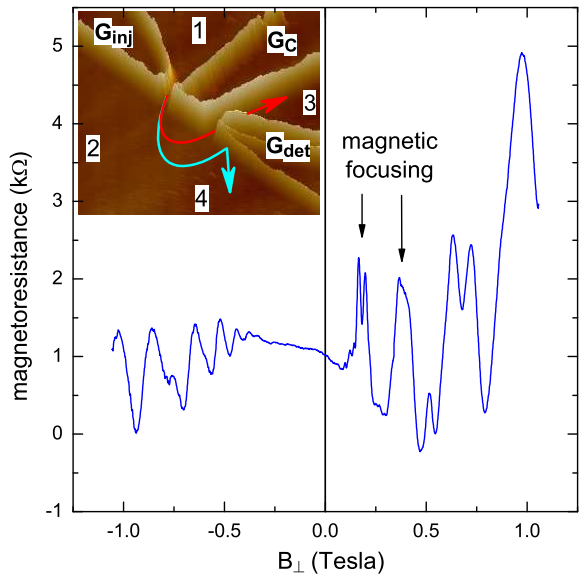} \vspace{-0.3in}
\caption{(color online) Magnetoresistance and layout of focusing
devices. Voltage across the detector (contacts 3 and 4) is
measured as a function of magnetic field perpendicular to the
surface of the sample ($B_{\bot}$). Lithographical separation
between point contacts is 0.8 $\mu$m. Current of 1 nA is flowing
through the injector (contacts 1 and 2). Positions of the magnetic
focusing peaks are marked with arrows. Inset: AFM micrograph of
sample A ($5\protect\mu$m$ \times5\protect\mu$m). Light lines are
the oxide which separates different regions of 2D hole gas.
Conductance of quantum point contacts is controlled via voltages
applied to the detector $G_{det}$, injector $G_{inj}$ and the
central $G_C$ gates. Semicircles show schematically the
trajectories for two spin orientations.}
\label{\ffile}
\end{figure}

Our approach is to use intrinsic spin-orbit (SO) interactions
existing in low-dimensional systems. It has long been appreciated
that such interactions can be interpreted as an effective
momentum-dependent magnetic field that influences spin of charge
carriers\cite{rashba60}. More recently, it has been
recognized\cite{aronov93,loss90,aharonov84,aleiner01a} that SO
interactions can be also viewed as an effective orbital magnetic
field with an opposite sign for different spin orientations. Now,
assume that two different magnetic fields $B_{\pm}=B_{\bot}\pm
B_{so}$ affect orbital motion of charge carriers with two distinct
spins. Here, $B_{\bot}$ is the perpendicular external magnetic
field and $B_{so}$ is the spin-orbital effective field
characteristic for cyclotron motion. Then, charge carriers move
along the cyclotron orbit with spin-dependent radius
$R_c^{\pm}=p_f/e(B_{\bot}\pm B_{so})$, where
$p_f=\sqrt{2m\epsilon_f}$ and $\epsilon_f$ are the Fermi momentum
and energy, and $e$ is the charge of carriers. A spin-dependent
$R_c$ has been observed in commensurability
oscillations\cite{lu98}. Our goal here is to use spin-dependence
of $R_c$ for spatial separation of carriers with distinct spins.
To do this, we use magnetic
focusing\cite{sharvin65,tsoi75,houten89} that we show to be
spin-dependent as a result of SO interactions. In the magnetic
focusing configuration, charge carriers are injected in the
two-dimensional gas through the injector quantum point contact
(QPC), propagate along the orbits defined by $R_c^{\pm}$, and are
detected by the detector QPC. By adjusting $R_c$, we select the
spin of charge carriers that reach the detector. Tuning $R_c$ is
possible by either changing $B_{\bot}$, changing electron density
with electrostatic gates ($p_f^2=2\pi\hbar^2 n$, where $n$ is the
carrier density), or changing $B_{so}$ by adjusting external
electric fields~\cite{miller02,papadakis02}.

To demonstrate spatial separation of spins experimentally we
fabricated several devices in the magnetic focusing geometry from
two dimensional hole gas (2DHG), see inset in Fig.~\ref{fig1}. The
structure is formed using atomic force microscopy local anodic
oxidation technique (AFM LAO)\cite{snow94,held97,rokhinson02}.
Oxide lines separate the 2DHG underneath by forming $\sim200$ mV
potential barriers. A specially designed heterostructure is grown
by MBE on [113]A GaAs. Despite very close proximity to the surface
(350\AA), the 2DHG has an exceptionally high mobility
$0.4\cdot10^{6}$ V$\cdot$s/cm$^2$ and relatively low hole density
$n=1.38\cdot10^{11}$ cm$^{-2}$. The device consists of two QPCs
separated by a central gate. Potential in the point contacts can
be controlled separately by two gates $G_{inj}$ and $G_{det}$, or
by the central gate $G_{C}$ . In our experiments the central gate
was kept at $-0.3$ V and $\sim0.2$ V were applied to the gates
$G_{inj}$ and $G_{det}$. Asymmetric biasing of point contacts
provides sharper confining potential and reduces the distance
between the two potential minima by $\Delta L\sim 0.07\ \mu$m.

\begin{figure}[t]
\def\ffile{fig2}
\centering\includegraphics[scale=1.3]{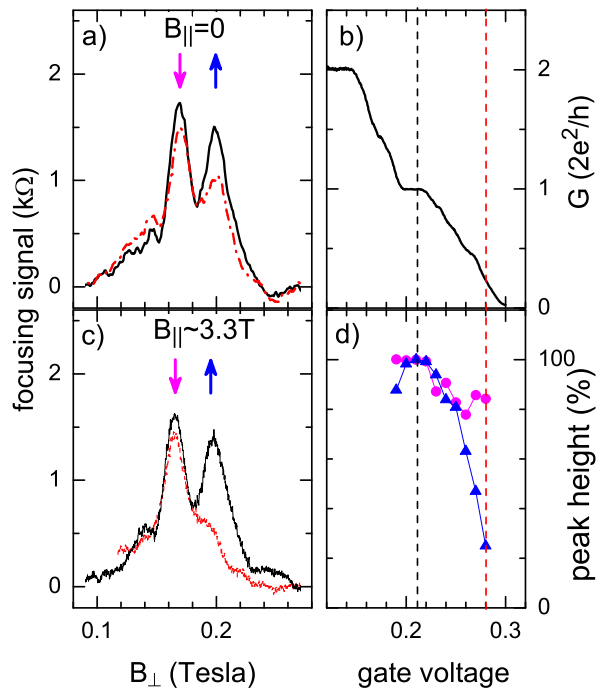} \vspace{-0.0in}
\caption{(color online) Magnetoresistance is measured with
magnetic filed oriented (a) perpendicular or (c) at
$\sim3^{\circ}$ to the surface. Series resistance due to 2DHG is
subtracted. In-plane field $B_{||}=3.3$ T corresponds to the
center of the peak and is aligned with the direction of the
carriers injection. Black curves are measured with injector and
detector QPCs gated to pass both spin orientations (conductance
$G=2e^2/h$). Red curves are measured with injector QPC gated at
$G\approx0.5e^2/h$. In (b) conductance of the injector QPC is
plotted as a function of the gate voltage at $B_{||}=3.3$ Tesla.
(d) Height of the low- (magenta) and high-field (blue) peaks in
(b) is plotted as a function of the injector gate voltage.}
\label{\ffile}
\end{figure}

Magnetic focusing manifests itself as equidistant peaks in
magnetoresistance $R(B_{\bot})$ for only one direction of
$B_{\bot}$. $R$ is measured by applying small current through the
injector QPC while monitoring voltage across the detector QPC, see
inset in Fig.~\ref{fig1}. At $B_{\bot}<0$ cyclotron motion forces
carriers away from the detector. Than, only 2DHG contributes to
$R$, which has almost no $B_{\bot}$-dependence at low fields and
shows Shubnikov--de Haas oscillations at $|B_{\bot}|>0.3$ T. For
$B_{\bot}>0$ several peaks due to magnetic focusing are observed.
Peaks separation $\Delta B\approx0.18$ T is consistent with the
expected value for the lithographical distance between the
injector and detector QPCs $L=0.8\ \mu$m. The data is symmetric
upon exchange of the injector and detector and simultaneous
reversal of the magnetic field direction.

The first focusing peak is a doublet consisting of two peaks
separated by $36$ mT. These peaks are the sharpest when both QPCs
are gated to pass exactly one spin-degenerate mode (within the
$G=2e^2/h=12.9$ k$\Omega$ conductance plateau in the QPC
characteristic, Fig.~\ref{fig2}b). Peaks in a doublet have
approximately the same height.

A QPC can be used as a spin filter for one spin polarization if
spin degeneracy is lifted by Zeeman splitting of energy levels in
a strong magnetic field\cite{potok02}. We apply an in-plane field
$B_{||}$, which has little effect on the cyclotron motion of holes
but acts on their spin degrees of freedom. At $B_{||}=3.3$ T there
is a pronounced step at $e^2/h$ in conductance vs gate voltage
characteristic in both QPCs, see Fig.~\ref{fig2}b. The appearance
of this step means that Zeeman energy exceeds both the broadening
of the transverse quantized energy levels of the QPC and the
temperature. Then, for $G\le e^2/h$ only holes with one spin
polarization are allowed to pass through the point contact.
Experiments\cite{potok02} with electrons had shown reduction of
the height of the focusing peaks by 50\% due to spin filtering,
when conductance of one QPC was tuned below $e^2/h$ and
conductance of the other QPC was maintained at $2e^2/h$.

We use the spin filtering by QPCs at high $B_{||}$ to probe the
spin states which correspond to the first focusing peak doublet.
As the conductance of the injector QPC is reduced below $2e^2/h$,
the height of the high-field peak in the doublet reduces while the
height of the low-field peak remains almost the same, see
Fig.~\ref{fig2}c,d. When conductance of the injector QPC is
$\sim0.5 e^2/h$, the high-field peak almost vanishes in a striking
contract to the electron case. Similarly, the high-field peak
vanishes if detector acts as a spin filter while injector is tuned
to accept both spin polarizations. In contrast, at zero $B_{||}$
peaks relative strength does not change significantly as the
injector conductance is decreased, see Fig.~\ref{fig2}a. A small
suppression of the high-field peak at $B_{||}$ can be attributed
to a partial lifting of the spin degeneracy in a non-zero focusing
field $B_{\bot}=0.2$ T. Therefore, we conclude that the two peaks
within the doublet correspond to the two spin states of the holes.
The peak spacing of 36 mT means that focusing points for the two
spin states are $\approx 120$ nm apart.

In order to explain the effect qualitatively, we assume that
charge carriers in GaAs quantum well are characterized by the
isotropic kinetic energy and the Dresselhaus spin-orbit
interaction, so that the Hamiltonian can be written as
\cite{altshuler81} $H=\frac{1}{2m}(p_x+\gamma\sigma_x)^2+
\frac{1}{2m} (p_y-\gamma\sigma_y)^2$, where $m$ is the effective
mass, $\vec{p}$ is the electron momentum, $\sigma_i$ are the Pauli
matrices ($i=x,y$), and $\gamma$ is the spin-orbit parameter. For
simplicity, we neglect anisotropy of the effective mass that do
not change the qualitative picture. In the semiclassical
description, appropriate for the range of magnetic fields
\textbf{$B_{\bot}$} used for the focusing, the motion is described
by simple equations
\begin{eqnarray}
&&\frac{d\vec{p}}{dt}=e \vec{v}\times \vec{B}\quad\quad
\vec{v}= \frac{d\vec{r}}{dt}= \frac{\partial \epsilon_{\pm}
(\vec{p})} {\partial\vec{p}}  \nonumber \\
&&\epsilon_{\pm}=\frac{1}{2m} (p\pm\gamma)^2+\frac{\gamma^2}{2m},
\label{eq0}
\end{eqnarray}
where $\vec{r}$, $\vec{v}$ and $\epsilon_{\pm}$ are the charge
carrier coordinate, velocity and energy for the two spin
projections. This description implies that carrier wavelength is
smaller than the cyclotron radius, and that jumps between orbits
with different spin projections are absent, i.e. $\epsilon_f\gg
\gamma p/m\gg \hbar\omega_c$. Eqs.~(\ref {eq0}) show that the
charge carrier with energy $\epsilon_{\pm}=\epsilon_f$ is
characterized by the spin-dependent trajectory: momentum
$\vec{p}_{\pm}$, coordinate $\vec{r}_{\pm}$ and cyclotron
frequency $\omega_c^{\pm}$.  The solution to these equations is
\begin{eqnarray}
&&p^{(x)}_{\pm}+ip^{(y)}_{\pm}= p_{\pm}\exp{(-i\omega^{\pm}_ct)}
\nonumber \\
&&r^{(x)}_{\pm}+ir^{(y)}_{\pm}= \frac{i \sqrt{2m\epsilon_f}}
{m\omega_c^{\pm}} \exp{(-i\omega^{\pm}_c t)}  \nonumber \\
&&\omega_c^{\pm}=\frac{eB_{\bot}}{m}(1\pm \gamma/p_{\pm}).
\label{eq2}
\end{eqnarray}
Thus, as discussed in the introduction, the cyclotron motion is
characterized by the spin-dependent field $B_{\pm}=B_{\bot}(1\pm
\gamma/p_{\pm})$. Using the semiclassical limit of the quantum
description, one obtains the identical
results\cite{bychkov84a}.

In the focusing configuration QPCs are used as monochromatic point
sources. Holes, injected in the direction perpendicular to the
two-dimensional hole gas boundary, can reach the detector directly
or after specular reflections from the boundary. As follows from
Eqs.~(\ref {eq2}), for each of the two spin projections there is a
characteristic magnetic field such that the point contact
separation is twice the cyclotron radius for a given spin, $L=2
R^c_{\pm}=2p_f/e B_{\pm}$, $p_f=\sqrt{2m\epsilon_f}$. The first
focusing peak occurs at
\begin{equation}
B_{\bot}^{\pm}=\frac{2(p_f\mp\gamma)}{eL}.
\label{eq3}
\end{equation}
The magnitude of $\gamma$ can be calculated directly from the peak
splitting $\gamma=(B_{\bot}^{+}-B_{\bot}^{-}) e L/4=7\cdot
10^{-9}$ eV$\cdot$s/m. A larger value of $\gamma\approx25\cdot
10^{-9}$ eV$\cdot$s/m was extracted from the splitting of
cyclotron resonance at 3 times higher hole
concentration\cite{stormer83}. For electrons, a much smaller value
$\gamma\approx1.5\cdot 10^{-9}$ eV$\cdot$s/m characterizes the
combined spin-cyclotron resonance\cite{stein83}. We note that
Eq.~(\ref{eq3}) is more general than the Eqs.~(\ref{eq2}). The
coefficient $\gamma$ essentially describes the separation in
momentum space of the two parts of the Fermi surface which
correspond to $\epsilon^{\pm}=\epsilon_f$, and includes
contributions of various spin-orbit terms in the 2D hole gas.
Analysis of the second and higher focusing peaks is complicated by
mixing of spin states due to reflections from the boundary. If
carriers necessarily undergo transitions between $\epsilon^+$ and
$\epsilon^-$ parts of the Fermi surface upon reflections, no
splitting of the second and other even focusing peaks is expected
\cite{usaj04}.

We now discuss spin states and their filtering by
QPCs in more detail. In the presence of SO interactions carriers
are characterized by the projection of their spin on the total
magnetic field, which is comprised of the external magnetic field
and the effective momentum-dependent SO field. In our devices, the
characteristic energy of SO interactions $\gamma p_f/m\approx0.2$
meV is larger than the Zeeman energy, and the quantum states of
the 2D holes are well characterized by the sign of the spin
projection onto their momentum $\vec{p}$ ($g^*\mu_BB\approx0.08$
meV at 3.5 T, $g^*\approx0.4$ for [$\bar{2}$33] crystallographic
direction). $R_c$ for these 2D states are different for different
spin projection signs.

However, in a QPC the hole momentum decreases and even vanishes on
the plateau, so that $\gamma p/m<g^*\mu_BB$. In this case holes
are characterized by their spin projection on the external
magnetic field. Thus, they can be spin-filtered by gating the QPC
to $G<e^2/h$. A hole, leaving injector in a certain spin state,
will enter detector with the same spin orientation as long as 1D
spin states in both QPCs adiabatically evolve into
$\vec{p}$-projection states in the 2DHG, and spin evolution is
adiabatic along the ballistic cyclotron trajectory.

The spin-dependent focusing field $B^{\pm}_{\bot}$ is proportional
to the spin-orbit constant $\gamma$ and does not depend on the
cyclotron frequency $\omega_c=eB_{\bot}/m$. At the same time, the
spin-dependent cyclotron frequency in Eq.~\ref{eq2} is
proportional to both $\omega_c$ and $\gamma$. Thus, the effective
magnetic field $B_{so}$ is itself proportional to $B_{\bot}$. This
effect differs from the spin-dependent shift of the Aharonov-Bohm
oscillations in the conductance of rings, where the additional
spin-orbit flux and the Aharonov-Bohm flux are independent of each
other\cite{aronov93}. If Zeeman effect is taken into account, both
$\omega_c^{\pm}$ and $R_c^{\pm}$ acquire additional dependence on
$B_{\bot}$, as well as on $B_{||}$. However, in the present
experimental setting the Zeeman splitting is small compared to the
effects of SO interactions and is essential for filtering spins in
the injector and detector QPCs only.

The phenomenon we report in this Letter is not restricted to holes
in GaAs but is generic to any system with intrinsic spin-orbit
interactions. We believe that exceptional quality two-dimensional
hole systems provides novel opportunities for spin manipulation.
In particular, it will be possible to realize other schemes of
spin separation and filtering\cite{datta90,khodas04}.

In conclusion, we developed a method to spatially separate spin
currents in materials with intrinsic spin-orbit interactions by a
weak magnetic field. This has been achieved in semiclassical
cyclotron motion, where distinct spins are characterized by
distinct cyclotron radii and frequencies. Thus the cyclotron
motion can separate not only the particles with different masses
but also particles with different spin-orbit parameters. We
confirm the spatial separation of spins experimentally by
selectively detecting spin-polarized currents.

Authors thank A.~M. Finkelstein and G.~F. Giuliani for
enlightening discussion. The work was partially funded by DARPA
and NSA/ARDA.

\bibliographystyle{revtex}
\bibliography{rohi,focusing}

\end{document}